\documentstyle[11pt,epsfig,aaspp4]{article}

\received{ }
\accepted{ }

\lefthead{Catanese et al.}
\righthead{$>$350 GeV Gamma Rays from 1ES 2344+514}

\begin{document}

\title{Discovery of $>$350 GeV Gamma Rays from the BL Lacertae Object 
1ES 2344+514}

\author{M. Catanese,\altaffilmark{1}
C. W. Akerlof,\altaffilmark{2}
H. M. Badran,\altaffilmark{3}
S. D. Biller,\altaffilmark{4}
I. H. Bond,\altaffilmark{5}
P. J. Boyle,\altaffilmark{6}
S. M. Bradbury,\altaffilmark{5}
J. H. Buckley,\altaffilmark{7}
A. M. Burdett,\altaffilmark{5}
J. Buss\'{o}ns Gordo,\altaffilmark{6}
D. A. Carter-Lewis,\altaffilmark{1}
M. F. Cawley,\altaffilmark{8}
V. Connaughton,\altaffilmark{9}
D. J. Fegan,\altaffilmark{6}
J. P. Finley,\altaffilmark{10}
J. A. Gaidos,\altaffilmark{10}
T. Hall,\altaffilmark{10}
A. M. Hillas,\altaffilmark{5}
F. Krennrich,\altaffilmark{1}
R. C. Lamb,\altaffilmark{11}
R. W. Lessard,\altaffilmark{10}
C. Masterson,\altaffilmark{6}
J. E. McEnery,\altaffilmark{6,13}
G. Mohanty,\altaffilmark{1,14}
J. Quinn,\altaffilmark{12}
A. J. Rodgers,\altaffilmark{5}
H. J. Rose,\altaffilmark{5}
F. W. Samuelson,\altaffilmark{1}
M. S. Schubnell,\altaffilmark{2}
G. H. Sembroski,\altaffilmark{10}
R. Srinivasan,\altaffilmark{10}
T. C. Weekes,\altaffilmark{12}
C. W. Wilson,\altaffilmark{10}
and J. Zweerink\altaffilmark{1}
}

\altaffiltext{1}{Department of Physics and Astronomy, Iowa State
University, Ames, IA 50011-3160}
\altaffiltext{2}{Randall Laboratory of Physics, University of Michigan,
Ann Arbor, MI 48109-1120}
\altaffiltext{3}{Department of Physics, Faculty of Science, Tanta University, 
Tanta, Egypt}
\altaffiltext{4}{Department of Physics, Oxford University, Oxford, UK}
\altaffiltext{5}{Department of Physics, University of Leeds,
Leeds, LS2 9JT, Yorkshire, England, UK}
\altaffiltext{6}{Experimental Physics Department, University College, 
Belfield, Dublin
4, Ireland}
\altaffiltext{7}{Department of Physics, Washington University, St. Louis,
MO 63130}
\altaffiltext{8}{Physics Department, St.Patrick's College,
Maynooth, County Kildare, Ireland}
\altaffiltext{9}{NASA, Marshall Space Flight Center, Huntsville, AL 35812}
\altaffiltext{10}{Department of Physics, Purdue University, West
Lafayette, IN 47907}
\altaffiltext{11}{Space Radiation Laboratory, California Institute of
Technology, Pasadena, CA 91125}
\altaffiltext{12}{ Fred Lawrence Whipple Observatory, Harvard-Smithsonian 
CfA, P.O. Box 97, Amado, AZ 85645-0097} 
\altaffiltext{13}{Present address: Department of Physics, University of Utah,
Salt Lake City, UT 84112}
\altaffiltext{14}{Present address: LPNHE \'{E}cole Polytechnique, 91128
Palaiseau CEDEX, France}

\authoremail{catanese@egret.sao.arizona.edu}
\authoraddr{Michael Catanese, F. L. Whipple Observatory, P.O. Box 97, 
Amado, AZ 85645}

\begin{abstract} 

We present the discovery of $>$350 GeV gamma-ray emission from the BL
Lacertae (BL Lac) object 1ES 2344+514 with the Whipple Observatory 10m
gamma-ray telescope.  This is the third BL Lac object detected at very
high energies (VHE, $E>300$ GeV), the other two being Markarian 421
(Mrk 421) and Mrk 501.  These three active galactic nuclei are all
X-ray selected and have the lowest known redshifts of any BL Lac
objects currently identified.  The evidence for emission from 1ES
2344+514 comes mostly from an apparent flare on 1995 December 20
(universal date) during which a 6$\sigma$ excess was detected with an
average flux of I($>$350 GeV) $ = 6.6 \pm 1.9 \times 10^{-11}$ photons
cm$^{-2}$ s$^{-1}$.  This is approximately 63\% of the VHE emission
from the Crab Nebula, the standard candle in this field.  Observations
taken between 1995 October and 1996 January, excluding the night of
the flare, yield a 4$\sigma$ detection indicating a flux level of
I($>$350 GeV) $ = 1.1 \pm 0.4 \times 10^{-11}$ photons cm$^{-2}$
s$^{-1}$, or about 11\% of the VHE Crab Nebula flux.  Observations
taken between 1996 September and 1997 January on this object did not
yield a significant detection of a steady flux, nor any evidence of
flaring activity.  The 99.9\% confidence level upper limit from these
observations is I($>$350 GeV) $ < 8.2 \times 10^{-12}$ photons
cm$^{-2}$ s$^{-1}$, $\lesssim$8\% of the Crab Nebula flux.  The low
baseline emission level and variation in nightly and yearly flux of
1ES 2344+514 are the same as the VHE emission characteristics of Mrk
421 and Mrk 501.

\end{abstract}


\setcounter{footnote}{0}
\section{Introduction}
\label{intro}

The Whipple collaboration has discovered the two extragalactic sources
of very high energy (VHE, E$\gtrsim$300 GeV) gamma rays Markarian 421
(\cite{Punch92}) and Markarian 501 (\cite{Quinn96}).  Both of these
are BL Lacertae (BL Lac) objects, a sub-class of the blazar class of
active galactic nucleus (AGN).  Blazars are the only class of AGN
detected above 100 MeV by the Energetic Gamma-Ray Experiment Telescope
(EGRET) on the {\sl Compton Gamma-Ray Observatory} ({\sl CGRO})
(\cite{Fichtel94}; \cite{Thompson95}; \cite{Thompson96}) and BL Lac
objects make up a significant fraction of the blazars detected by
EGRET.  The most striking feature of the VHE emission from Markarian
(Mrk) 421 and Mrk 501 is the day-scale or shorter flaring which has
produced emission as high as 10 times the Crab flux (\cite{Gaidos96}),
but flares on the order of 1/2 the Crab flux are more typical
(\cite{Buckley96}; \cite{Quinn96}; \cite{Quinn97}; \cite{McEnery97}).
The baseline emission level for these two objects can be very low; Mrk
501 was initially detected to have a VHE flux 8\% that of the
Crab Nebula (\cite{Quinn96}) and Mrk 421 has dropped below the
detection limit of the Whipple Observatory gamma-ray telescope for as
long as a month (\cite{Buckley96}; \cite{McEnery97}).

Initial searches for extragalactic sources of VHE gamma rays by the
Whipple collaboration concentrated mainly on blazars detected by EGRET
or AGN of any type with small redshift (\cite{Kerrick95}).  This
approach led to the detection of Mrk 421, but yielded only upper
limits for the other objects studied.  Beginning in the spring of
1995, the Whipple collaboration initiated a more focussed observation
campaign on nearby BL Lac objects.  The search list was mainly drawn
from the work of Perlman et al. (1996), who identified BL Lac objects
from the Einstein Slew Survey sources (\cite{Elvis92}).\footnote{We
note that Mrk 501 was chosen as a candidate independently because of
its similarities to Mrk 421.} This list contained all of the prominent
BL Lac objects (e.g., BL Lacertae, PKS 2155-304, Mrk 421), but also
significantly increased the number of BL Lac objects with known
redshifts.  In addition, because these objects were drawn from an
X-ray survey, they were more likely to have emission spectra similar
to Mrk 421 and Mrk 501.  We limited the search to those sources with
low redshifts (initially $z < 0.1$, but eventually extended to $z <
0.2$) in order to reduce the chances that the VHE emission would be
significantly attenuated by interaction with the background
intergalactic infrared radiation fields (\cite{Gould67};
\cite{Stecker92}).  The first success of this observing program was
the detection of Mrk 501 as a gamma-ray source.  As Mrk 501 was not
identified in EGRET catalogs as a significant source, it highlighted
the ability of ground-based gamma-ray telescopes to not only
complement the results of the space-based gamma-ray telescopes, but to
augment them.  We report here on the second source detected in this BL
Lac object search program, 1ES 2344+514.

1ES 2344+514 was only recently identified as a BL Lac object
(\cite{Perlman96}), based on its lack of optical emission lines with
observed equivalent width greater than 5\AA \ and its Ca II ``break
strength" being smaller than 25\%.  The latter criteria is indicative
of the presence of a power law continuum and the former eliminates
quasars.  Perlman et al. (1996) determined this object to have a
redshift of $z = 0.044$ based on absorption lines; it had no evident
emission lines.  This makes it the third closest known BL Lac object, after
Mrk 421 and Mrk 501.  Perlman et al.  (1996) derive a 2 keV X-ray flux
of 1.14 $\mu$Jy, roughly 1/3 the flux detected for Mrk 421 and Mrk 501
in that same work, and measured an optical magnitude of $m_V = 15.5$
with no galaxy light subtraction.  Measurements taken with the Very
Large Array radio interferometer indicate that its radio emission is
``point-like," with more than 80\% of its flux being from an
unresolved point source (\cite{Patnaik92}; \cite{Perlman96}).  The
Green Bank radio survey lists its 5 GHz radio flux as 231$\pm$25 mJy,
which is about 1/3 and 1/4 the 5 GHz flux of Mrk 421 and Mrk 501,
respectively.

In the following sections we outline the observation and analysis
techniques (\S~\ref{analysis}), detail the observations taken by the
Whipple collaboration on 1ES 2344+514 between 1995 October and 1997
January (\S~\ref{obs}), and finally discuss the outcome and
implications of those observations (\S~\ref{discuss}).

\section{Observation and Analysis Techniques}
\label{analysis}

The VHE observations reported in this paper were made with the
atmospheric Cherenkov imaging technique (\cite{Cawley95}; \cite{Reynolds93}) 
using the
10~m optical reflector located at the Whipple Observatory on Mt.
Hopkins in Arizona (elevation 2.3 km) (\cite{Cawley90}).  A camera,
consisting of photomultiplier tubes (PMTs) mounted in the focal plane
of the reflector, records images of atmospheric Cherenkov radiation
from air showers produced by gamma rays and cosmic rays.  For most of
the observations reported here, the camera consisted of 109 PMTs (each
viewing a circular field of 0\fdg259 radius) with a total field of
view (FOV) of 3$^\circ$.  In 1996 December, 42 additional PMTs were
added to the camera, increasing the field of view to 3$\fdg$3.
Because only a small fraction of the data presented here was taken
with more than 109 PMTs, the analysis of this data only uses the
original 109 PMTs.  This makes the results consistent with the rest of
the 1996/97 season and allows one set of parameter cuts to be used in
the analysis.  The telescope is sensitive to gamma rays with energies
between approximately 200 GeV and 20 TeV.

We characterize the Cherenkov images' roughly elliptical shapes and
locations and orientations within the telescope FOV using a moment
analysis (\cite{Reynolds93}).  The gamma-ray selection applied here
utilizes the Supercuts95 criteria (Table 1; cf.  \cite{Quinn96};
\cite{Catanese96}), which were optimized on contemporaneous Crab
Nebula data to give the best sensitivity to point sources.  Monte
Carlo simulations indicate that this analysis results in an energy
threshold of $\sim$350 GeV and an effective area of $\sim 3.5 \times
10^8$ cm$^2$.  Details of the methods used to estimate the energy
threshold and effective area are given elsewhere (\cite{Mohanty97}.
The large effective area makes this technique very sensitive to
short-term variability in sources.

\subsection{On/Off Observations}
\label{onoff}

The traditional mode of observing potential sources with the Whipple
Observatory gamma-ray telescope, and the method which is still usually
used to confirm source signals and derive source spectra is the On/Off
mode of observation.  In this type of observation, the putative source
is tracked continuously for 28 minutes with the center of the
telescope FOV positioned at the source location.  In a 28 minute
control, or ``Off," run the telescope tracks a position offset by 30
minutes in right ascension but with the same declination as the
putative source.  This Off-source run begins exactly 30 sidereal
minutes before or after the start of the On-source run so that the
telescope tracks the same range of elevation and azimuth for both
runs.  Night sky brightness differences between the two observing
fields are equalized off-line by software padding (\cite{Cawley93}).
The significance of any excesses or deficits in the observations are
calculated using the maximum likelihood method of Li \& Ma (1983).

This data collection technique has been shown to produce statistically
stable gamma-ray count rates from the Crab Nebula over the course of
several months (\cite{Quinnt97}), and to allow consistent,
reproducible spectra to be derived for the existing sources
(\cite{Mohanty97}; \cite{Hillas97}; \cite{Zweerink97}).  However, this
method requires excellent weather conditions to obtain consistent
results between the On-source and Off-source runs so it reduces the
observing time available.  Also, because it only uses a small fraction
of its gamma-ray like events to estimate the background, it does not
give the most statistically accurate background estimates possible.

\subsection{Tracking Observations}
\label{track}

\begin{figure}
\centerline{\epsfig{file=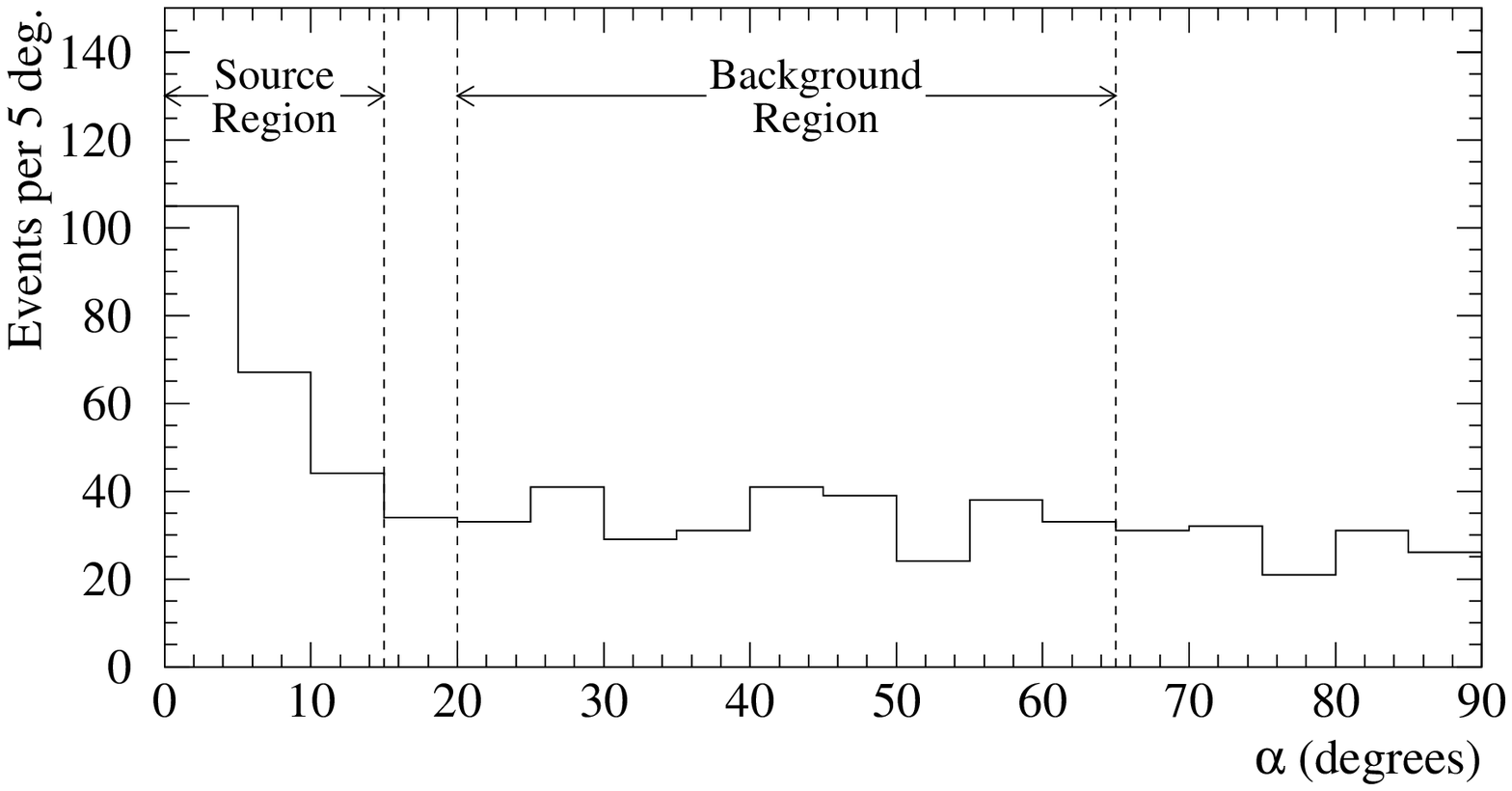,height=3in,angle=0.}}
\caption{The distribution of $\alpha$ parameter values for events
passing all but the $\alpha$ parameter cut of Supercuts95.  
These data were taken on
the Crab Nebula between 1996 October and 1997 January.  The source and
background regions used in the Tracking analysis described in the text
are indicated.  
\label{trkfig} 
}
\end{figure}

For the reasons listed above, the more frequent mode of observation,
particularly when searching for new sources, is the Tracking mode.
This mode of data collection for the Whipple Observatory gamma-ray
telescope has been described previously (\cite{Kerrick95};
\cite{Quinn96}), but we do so again here for clarity and because some
differences in the method of significance calculation are used.  In
this mode, only the On-source position is tracked, in runs of 28
minute duration, and no control observations are taken.  To estimate
the expected background, we use those events which pass all of the
Supercuts95 criteria except orientation (characterized by the $\alpha$
parameter).  We use events with values of $\alpha$ between 20$^\circ$
and 65$^\circ$ as the background region, as indicated in
Figure~\ref{trkfig}.  The lower bound gives a 5$^\circ$ buffer between
the On-source region and the Off-source region, to allow some room for
spillover events from the source region.  The upper bound is chosen
because the region beyond 65$^\circ$ is less stable from run to run
and source to source due to edge of field effects.  As shown in
Appendix~\ref{trkmode}, this non-standard method of background
estimation gives reliable results for a variety of observation fields.
Using contemporaneous non-source data, we find that the factor (called
a tracking ratio) which converts the number of events in the region
$\alpha = 20^\circ - 65^\circ$ to an estimate of the On-source
background ($\alpha = 0^\circ - 15^\circ$) is $r$ = 0.292$\pm$0.004
for the 1995/96 season and $r$ = 0.316$\pm$0.004 for the 1996/97
season.

To convert On-source and background counts to a significance, $S$, in
the tracking analysis, we use a simple propagation of errors formula:
\begin{equation} 
S = {{N_{on} - r*N_{bkd}}\over{\sqrt{N_{on} +
r^2*N_{bkd} + N_{bkd}^2*\Delta r^2}}} 
\label{signif} 
\end{equation}
where $N_{on}$ is the number of events in the source region ($\alpha <
15^\circ$), $N_{bkd}$ is the number of events in the background region
($\alpha = 20^\circ - 65^\circ$), and $r\pm\Delta r$ is the tracking
ratio and its statistical uncertainty.  The maximum likelihood method
of Li \& Ma (1983) cannot be used when the ratio for converting the
background counts to a background estimate has any uncertainty.  Li \&
Ma (1983) show that the estimate of the significance from equation
(\ref{signif}) tends to be conservative when $r < 1$.

\subsection{Flux and flux upper limit estimation}

After Supercuts95 are applied to an On/Off or Tracking data set, we
obtain a statistical significance as calculated with the methods
described in \S~\ref{onoff} and \S~\ref{track}, respectively, and a
corresponding gamma-ray count rate in terms of counts per minute.  If
the excess is not statistically significant, we calculate a 99.9\%
confidence level (CL) upper limit on the count rate using the method
of Helene (1983).  To convert these count rates or upper limits to
fluxes, we first express them as a fraction of the Crab Nebula count
rate for the same season.  This corrects for season to season
variations which affect the telescope response, and therefore its
gamma-ray count rate.  For the 1995/96 season, the gamma-ray count
rate for the Crab Nebula with a Supercuts95 analysis is
1.58$\pm$0.05/min.  In 1996/97, the Crab Nebula rate was
1.69$\pm$0.07/min.  Analysis of the Crab Nebula data shows that for
runs taken under good weather conditions, the gamma-ray count rate
does not change significantly within a season but can change
significantly between seasons (\cite{Quinnt97}.  So, we can assume the
gamma-ray count rate for a source can be reliably compared with that
of the Crab Nebula over the course of a season.

Once we have the flux or flux limit expressed as a fraction of the
Crab Nebula count rate, we multiply it by the Crab Nebula flux (in
units of photons cm$^{-2}$ s$^{-1}$) above the energy threshold of the
observations.  We determined the energy threshold for the observations
reported here to be 350 GeV.  The integral Crab Nebula photon flux is
$I(>350 \ {\rm GeV})=(1.05 \pm 0.24) \times 10^{-10}$ photons
cm$^{-2}$ s$^{-1}$ (\cite{Hillas97}).  For flux estimates, we
propagate the uncertainty in the Crab Nebula gamma-ray count rate and
flux through to the final flux estimate, so the flux uncertainties
have a significant contribution from the Crab Nebula flux uncertainty.
The significances of an excess should thus not be estimated by the
photon flux uncertainties.  For the flux upper limits, we do not
propagate through the Crab Nebula uncertainties.  Upper limits are an
estimate of the flux that could be present in the data set but not
produce a significant excess.  This is most accurately derived from
the count rate because that is what determines the statistical
significance of the excess.  The Crab Nebula count rate and flux
uncertainties affect only the normalization, so the flux upper limits
quoted in terms of photons cm$^{-2}$ s$^{-1}$ have an uncertainty of
$\sim$25\%, mainly from the uncertainty in the Crab Nebula photon
flux.

This flux calculation takes advantage of the fact that the VHE Crab
Nebula flux is steady over at least a 7 year period (\cite{Hillas97})
so that changes in the Crab Nebula count rate are most likely due to
changes in the telescope sensitivity or threshold.  The one
disadvantage of this method is that it assumes the spectrum of the
putative source is similar to that of the Crab Nebula ($\propto
E^{-2.5}$).  Preliminary estimates of the spectrum for 1ES 2344+514
(to be presented in future work) indicate that it is not significantly
different than this.

As shown below, 1ES 2344+514 is observed with an average elevation of
$\sim 55^\circ$, much lower than the average elevation of the Crab
Nebula observations ($\sim 72^\circ$).  While the telescope trigger
rate does vary significantly with elevation, Supercuts95 analysis of
the Crab Nebula indicates no significant variation in the gamma-ray
count rate with elevation down to 50$^\circ$.  So, normalizing to the
Crab Nebula count rate should be valid over that elevation range.

\section{Observations and Results}
\label{obs}

For the 1995/96 observing season, 1ES 2344+514 was observed between
October and January.  After filtering the runs for bad weather and
instrumental problems, the data set consists of 19 Tracking runs
(exposure = 9.0 hours) and 26 On/Off pairs (exposure = 11.5 hours).
The elevation of these runs varied from 37$^\circ$ to 70$^\circ$ with
a mean of 55$^\circ$.  For the On/Off pairs, using the Supercuts95
gamma-ray selection criteria, we obtain a significance of 2.9$\sigma$
with a corresponding excess count rate of 0.19$\pm$0.07/min or a
99.9\% CL upper limit of $<$0.40/min.  This converts to a flux
upper limit of $<$0.25 times the VHE Crab Nebula flux rate or an
integral flux upper limit of I($>$350 GeV) $< 2.6 \times 10^{-11}$
photons cm$^{-2}$ s$^{-1}$.  

\begin{figure}
\centerline{\epsfig{file=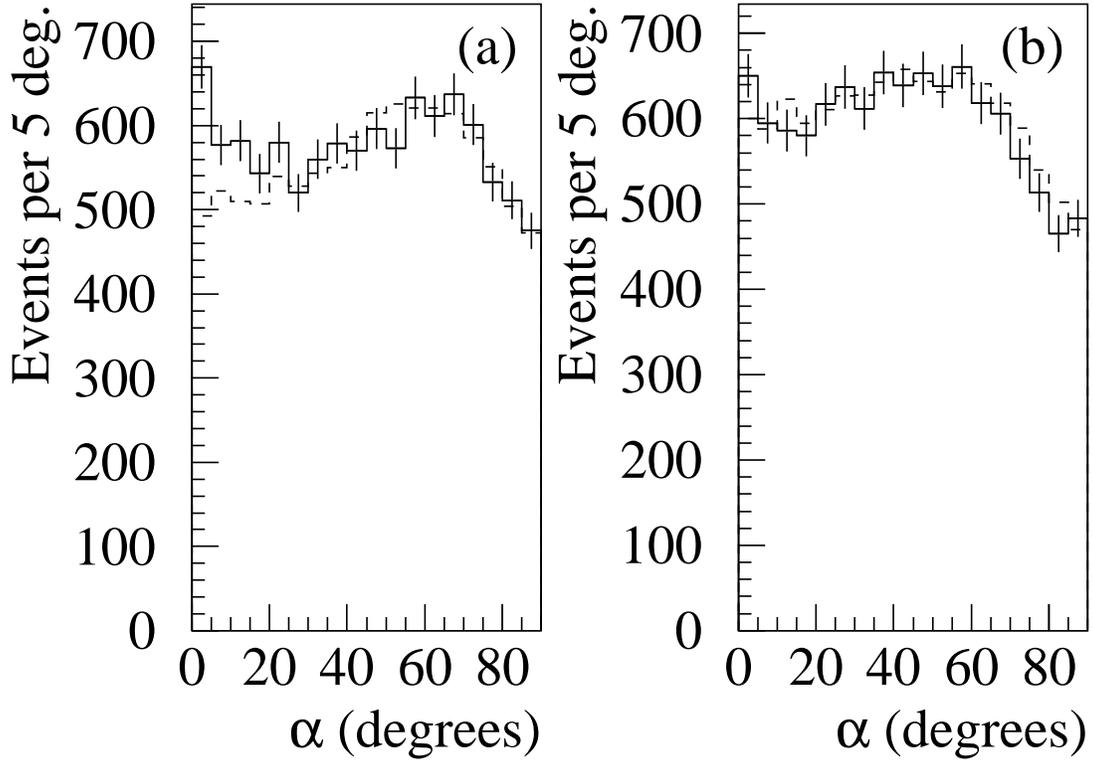,height=4in,angle=0.}}
\caption{Distributions of the $\alpha$ parameter for 1ES 2344+514 for
events which pass all other Supercuts95 cuts.  Shown are 
all On-source data taken (a) in 1995/96 and (b) in 1996/97.
The dashed curves are the $\alpha$ distributions for the non-source runs
used in each season 
to estimate a tracking ratio (see text for details) normalized
to the predicted background in the $\alpha = 0^\circ - 15^\circ$ region.
\label{alphatot}
}
\end{figure}

Combining all of the On-source runs whether taken as On/Off pairs or
Tracking runs, and using the Tracking analysis described in
\S~\ref{track}, we obtain a significance of 5.8$\sigma$ and count rate
of 0.25$\pm$0.04/min.  This translates into a flux of 0.16$\pm$0.03
times the Crab Nebula flux or I($>$ 350 GeV) = $1.7 \pm 0.5 \times
10^{-11}$ photons cm$^{-2}$ s$^{-1}$.  The higher significance is due
to a combination of having more data and the tracking analysis having
a more statistically accurate estimate of the background.  The count
rates for the two data sets are consistent within statistical errors.
These results are summarized in Table~\ref{data}.
Figure~\ref{alphatot} shows the $\alpha$ distribution for the summed
On-source runs, along with the $\alpha$ distribution for the
non-source runs used to estimate the tracking ratio.  The latter
distribution is normalized to the predicted background level from the
Tracking analysis.

For the 1996/97 observing season, 1ES 2344+514 was observed between
September and January.  After filtering for bad weather and
instrumental problems, the data consist of 38 On/Off pairs (exposure =
17.4 hours) and 16 Tracking runs (exposure = 7.5 hours).  The runs
were taken at elevations ranging from 46$^\circ$ to 70$^\circ$ with an
average elevation of 64$^\circ$.  For the On/Off pairs, Supercuts95
analysis resulted in a -0.4$\sigma$ deficit, with corresponding 99.9\%
CL upper limit $<$0.15 counts/min.  This is equivalent to an upper
limit on the flux of $<$0.09 times the Crab Nebula flux or I($>$350
GeV) $< 0.93 \times 10^{-11}$ photons cm$^{-2}$ s$^{-1}$.  For the
combination of the On-source runs taken as part of On/Off pairs or as
Tracking runs, Supercuts95 analysis gives a 0.4$\sigma$ excess with
corresponding 99.9\% CL upper limit $<$0.13 counts/min.  This upper
limit is equivalent to $<$0.08 times the VHE Crab Nebula flux or
I($>$350 GeV) $< 0.82 \times 10^{-11}$ photons cm$^{-2}$ s$^{-1}$.
These results are summarized in Table~\ref{data}.  The $\alpha$
parameter distribution for the Tracking analysis is shown in
Figure~\ref{alphatot}.

\subsection{Variability}
\label{var}

\begin{figure}
\centerline{\epsfig{file=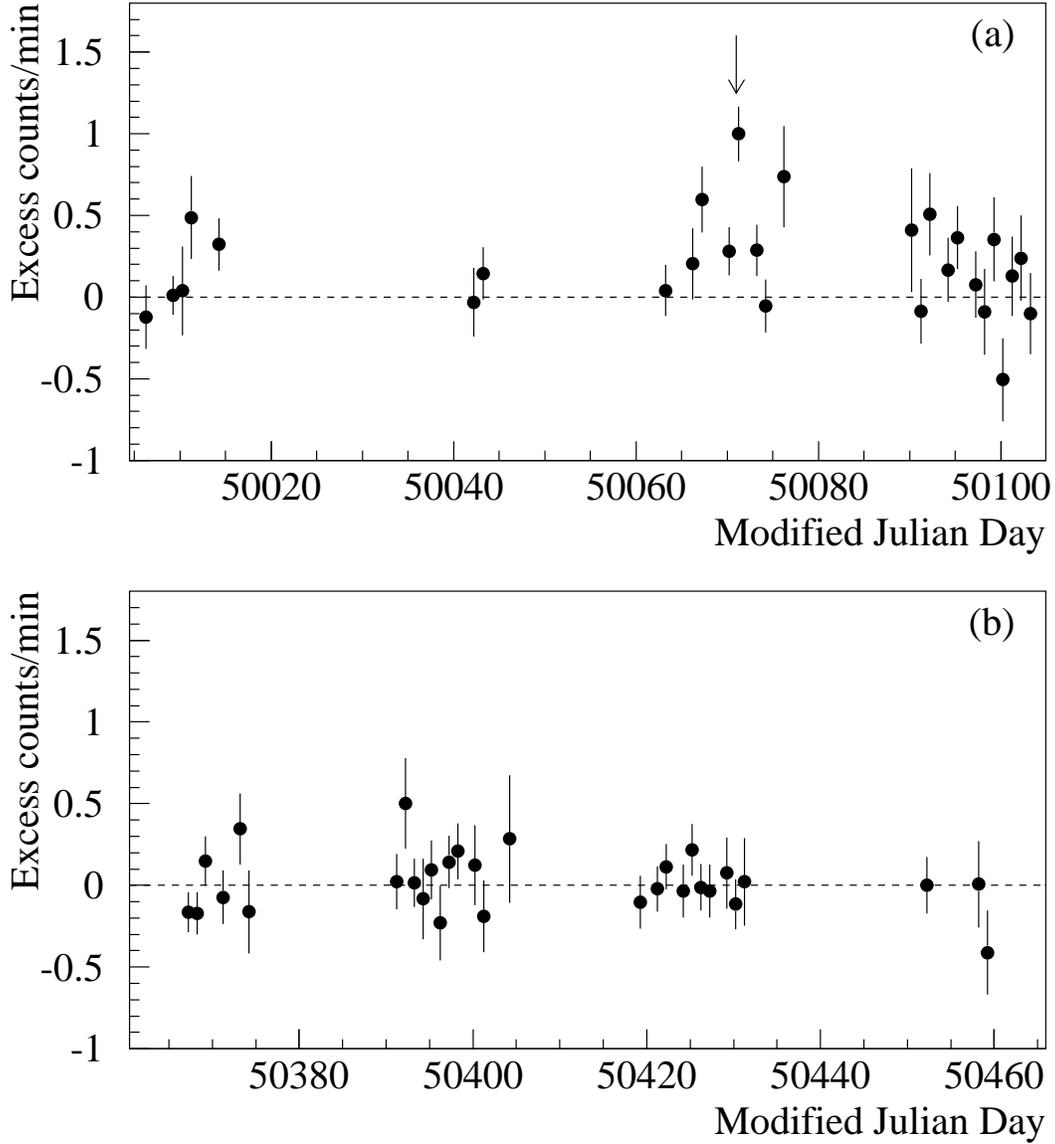,height=6in,angle=0.}}
\caption{Light curves for all On-source data taken in 
(a) 1995/96 and (b) 1996/97, for 
1ES 2344+514 passing Supercuts95.  The uncertainties are
1$\sigma$ statistical errors.  
The arrow in (a) indicates the flare on 1995 December 20 (MJD 50071).
In (b), MJD 50400 is 1996 November 13.
\label{lc}
}
\end{figure}

As discussed in \S~\ref{intro}, both Mrk 421 and Mrk 501 exhibit low
baseline levels of VHE emission, but with comparatively high flux
amplitude flares, typically lasting on the order of 1 day.  So, this
sort of behavior should be expected in the VHE emission from other BL
Lac objects if they emit VHE gamma rays.  Therefore, we have searched for
flaring activity from 1ES 2344+514 in both the 1995/96 and 1996/97
observing seasons.  The light curves showing the daily excess or
deficit for all On-source data taken during the two seasons are shown
in Figure~\ref{lc}.  For 1996/97, the observations show no evidence of
significant variability.  The maximum daily excess has a significance
of 1.8$\sigma$ and the $\chi^2$ probability that the excesses and
deficits are consistent with statistical fluctuations about a common
mean is 0.80.  In contrast, for
the 1995/96 season the $\chi^2$ probability that the observations are
consistent with constant emission is $3.5\times 10^{-8}$.  For
comparison, the same analysis of the Crab Nebula data gives a $\chi^2$
probability of constant emission of 0.78 in 1995/96 and 0.94 in
1996/97.

\begin{figure}
\centerline{\epsfig{file=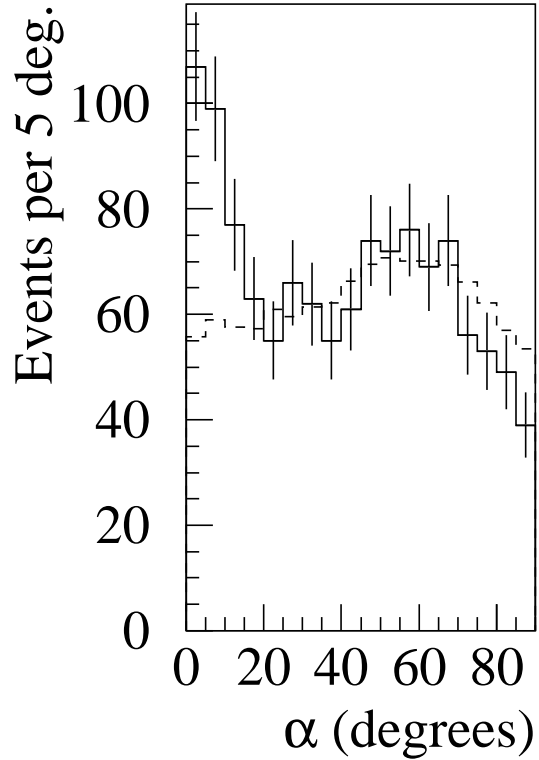,height=4in,angle=0.}}
\caption{Distribution of the $\alpha$ parameter for all On-source data
taken on 1ES 2344+514 on 1995 December 20 for
events which pass all other Supercuts95 cuts.  On/Off data
(a), and all On-source data (b) are shown.  The 
dashed curve is the $\alpha$ distribution for the non-source runs
used to estimate the 1995/96 tracking ratio (see text for details) normalized
to the predicted background in the $\alpha = 0^\circ - 15^\circ$ region.
\label{alpflr}
}
\end{figure}

\begin{figure}
\centerline{\epsfig{file=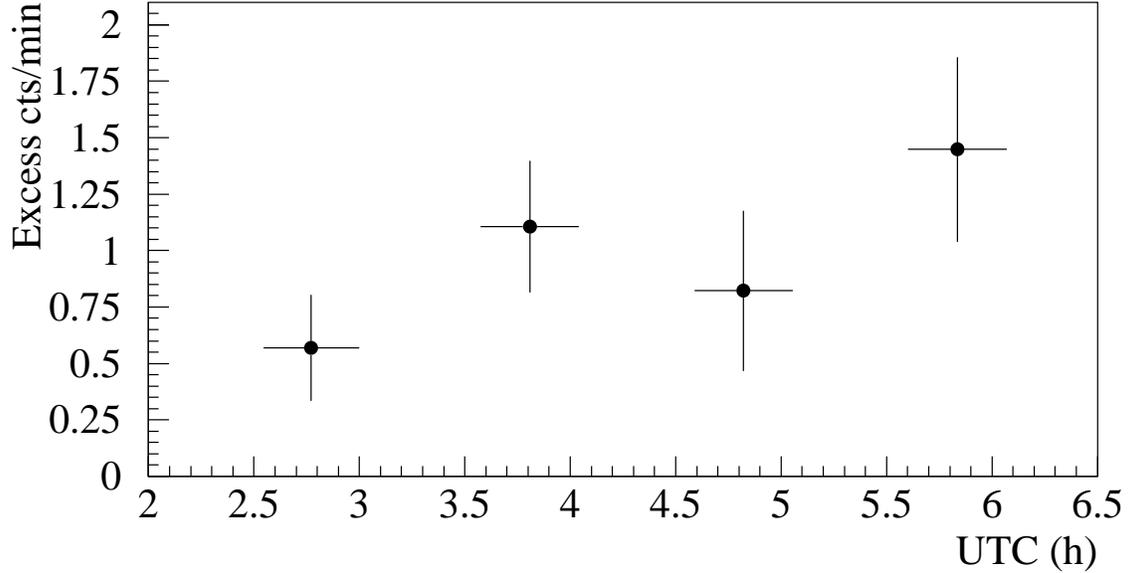,height=3in,angle=0.}}
\caption{The light curve for all On-source observations of 1ES 2344+514
taken on 1995 December 20 for events passing Supercuts95.  The error
bars are 1$\sigma$ statistical uncertainties.
\label{lc1220}
}
\end{figure}

The most significant contribution to the variability in 1995/96 is
1995 December 20 (indicated by the arrow in Figure~\ref{lc}).  On that
night, 3 On/Off pairs and 1 Tracking run were taken on 1ES 2344+514.
The On/Off pairs indicate a 4.2$\sigma$ excess and a flux of
0.55$\pm$0.13 times the Crab Nebula flux, or I($>$350 GeV) = $(5.8 \pm
1.9)\times 10^{-11}$ photons cm$^{-2}$ s$^{-1}$.  A Tracking analysis
of all 4 On-source runs reveals a 6.0$\sigma$ excess with a
corresponding flux of 0.63$\pm$0.11 times the Crab Nebula flux, or
I($>$350 GeV) = $(6.6 \pm 1.9) \times 10^{-11}$ photons cm$^{-2}$
s$^{-1}$.  The increase in significance for the Tracking analysis is
due to the additional run used.  The fluxes for the On/Off and
Tracking analyses are consistent within errors.  The $\alpha$
parameter distribution for the night of the flare is shown in
Figure~\ref{alpflr}, clearly indicating the excess.  There is no
statistically significant variation in the flux within the night (see
Figure~\ref{lc1220}).

The flare on 1995 December 20 constitutes slightly over 1/3 of the
excess gamma rays detected from 1ES 2344+514 in the 1995/96 season.
When the data from that night are removed, the Tracking analysis of
all the remaining On-source runs reveals an excess of 4.0$\sigma$ and
corresponding flux of 0.11$\pm$0.3 times the Crab Nebula flux, or
I($>$350 GeV) = $(1.1 \pm 0.4) \times 10^{-11}$ photons cm$^{-2}$
s$^{-1}$.  Thus, without the flare night, the evidence for emission
from 1ES 2344+514 is marginal.  The $\chi^2$ probability that the
emission is constant about the mean for the non-flare nights is 0.003,
indicating evidence for further variability at the 3$\sigma$ level.
>From Figure~\ref{lc} it is evident that any additional flaring was not
significant on day scales.  Given the low significance of the overall
excess, further study of this possible variability is not feasible.

\section{Discussion} 
\label{discuss}

We have presented evidence of VHE emission from the BL Lac object 1ES
2344+514, as detected with the Whipple Observatory gamma-ray
telescope.  The emission is detectable only in 1995/96 and only with
unambiguous statistical significance on 1995 December 20.  On the
latter date, the flux was $\sim$60\% that of the Crab Nebula.  The
baseline emission appeared to be $\sim$10\% of the Crab Nebula flux
during 1995/96 and was $<$8\% in 1996/97.  This low emission level
with a high flux flare is exactly what would be expected from a
VHE-emitting BL Lac object, based on what has been observed of the VHE
emission from the other known extragalactic VHE sources, Mrk 421 and
Mrk 501.  In fact, the initial detection of Mrk 501 as reported by
Quinn et al. (1996) is almost completely analogous to the observations
reported here for 1ES 2344+514.  The initial detection of Mrk 501
indicated a mean flux of 8\% that of the Crab Nebula and in more than
66 hours of observations, only a single flare was detected, with a
flux equivalent to 50\% of the Crab Nebula flux.  The one difference
is that in subsequent seasons, Mrk 501 has become brighter
(\cite{Quinn97}).  If it had become dimmer, it would not have been
detected in subsequent seasons, just as happened with 1ES~2344+514 in
1996/97.

The detection of 1ES 2344+514 is also consistent with Mrk 421 and Mrk
501 in that all three are nearby, X-ray selected BL Lac objects
(XBLs).  By XBLs we mean here those BL Lac objects which have
synchrotron spectra that extend into the UV to X-ray energy range.
This extension makes them bright soft X-ray sources, hence the XBL
label.  The low redshift means its VHE emission will not be attenuated
severely in the $\sim$300 GeV energy range by interaction with
background IR photons (\cite{Stecker96}; \cite{Biller97}).  BL Lac
objects have been suggested as better candidates for VHE emission than
the flat spectrum radio quasars (FSRQs) typically detected by EGRET
(\cite{Dermer94}) because BL Lac objects lack broad emission lines in
their optical spectra.  The absence of such lines may indicate less
VHE gamma-ray absorbing material at the source.  XBLs should be the
best blazar candidates for VHE emission if the gamma rays are produced
from inverse Compton scattering of low energy photons by the same
electrons which produce the synchrotron emission that dominates blazar
spectra from radio to optical or X-ray wavelengths (e.g.,
\cite{Konigl81}; \cite{Maraschi92}; \cite{Bloom96}; \cite{Sikora94};
\cite{Dermer92}).  This is because the synchrotron spectrum of XBLs
extends to higher energies than other blazars, like radio-selected BL
Lac objects (RBLs) and FSRQs which have synchrotron cutoffs at optical
wavelengths.  This implies the presence of higher energy electrons in
the jets of XBLs, so given similar magnetic fields and Doppler
factors, XBLs should have an inverse Compton spectrum which extends to
higher energies than those of RBLs or FSRQs (Sikora et al. 1994).
This picture is consistent with the lack of VHE emission from nearby
RBLs like BL Lacertae (\cite{Catanese97a}) and W Comae
(\cite{Catanese97b}) which are EGRET sources (\cite{Catanese97a};
\cite{Thompson96}).  However, it does not rule out models which
produce gamma-ray emission as the product of extremely high energy
protons in the AGN jet (e.g., \cite{Mannheim93}).  More detections of
BL Lac objects will be needed to further investigate that issue.

\begin{figure}
\centerline{\epsfig{file=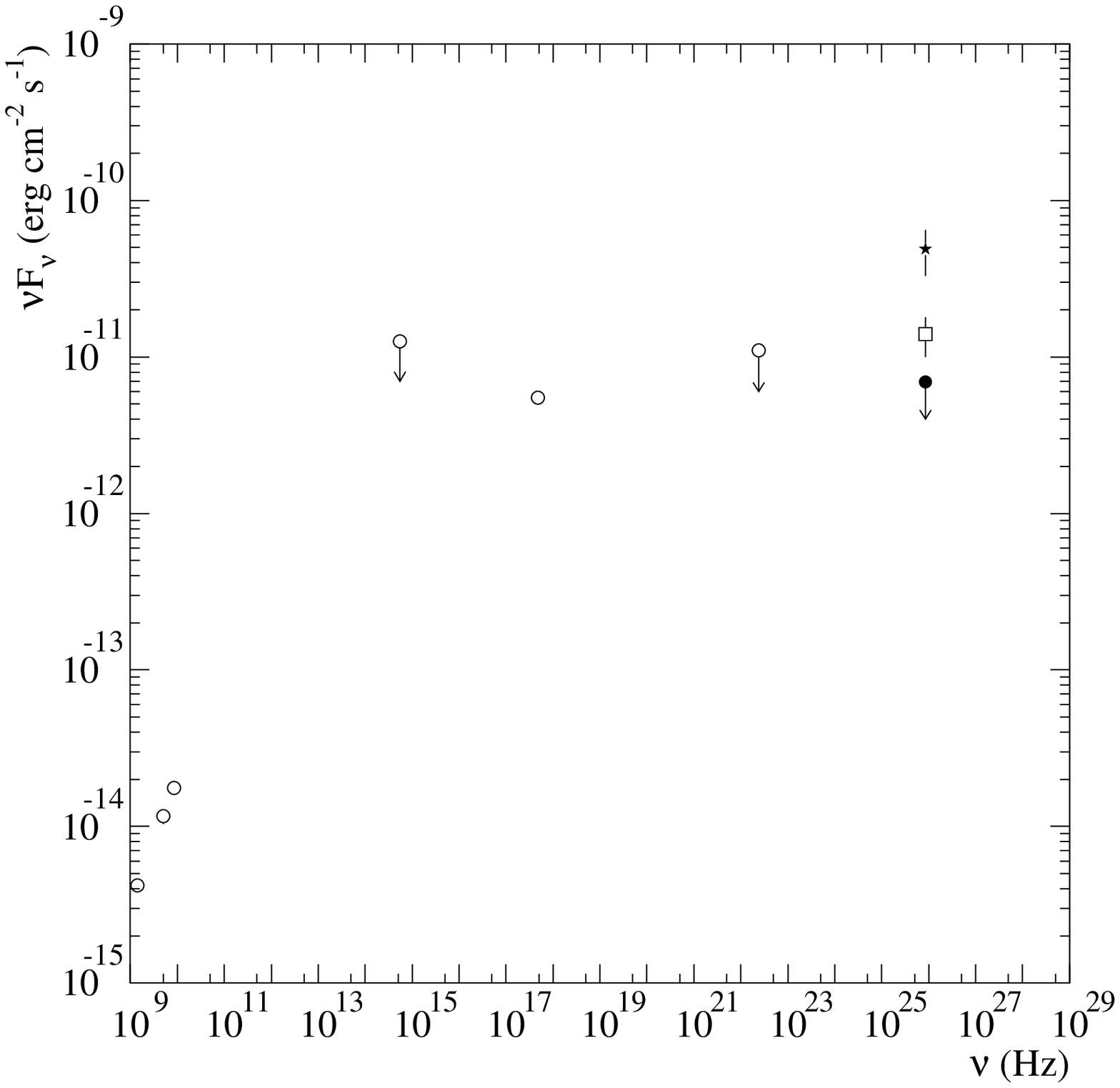,height=6in,angle=0.}}
\caption{The spectral energy distribution of 1ES 2344+514.  The VHE
observations from 1995/96 (open square), 1996/97 (filled circle), and
the flare flux of 1995 December 20 (filled star) are shown.
Non-contemporaneous X-ray, optical and radio fluxes are indicated by
the open circles (\protect\cite{Perlman96}; \protect\cite{Patnaik92};
\protect\cite{Gregory91}).  The optical flux is shown as an upper
limit because the galaxy light contribution is not subtracted.  In
addition, the EGRET upper limit is shown (D.J. Thompson 1996, for the 
EGRET team, private communication).
\label{nufnu} 
}
\end{figure}

The spectral energy distribution (SED) of 1ES 2344+514 is indicated in
Figure~\ref{nufnu}.  Because the data are not contemporaneous,
deriving limits on the jet parameters or making strong conclusions
about the shape of the SED are not feasible.  However, some general
conclusions can be drawn.  1ES 2344+514 is dimmer at every wavelength
than Mrk 421 and Mrk 501 and so it is perhaps not surprising that the
VHE emission of 1ES 2344+514 appears to be weaker on average.  If the
X-ray power output were similar to that at 350 GeV, as it is in Mrk
421 (\cite{Macomb95}; \cite{Buckley96}), the gamma-ray flux detected
in the flare would require an X-ray flux at 2 keV of 12$\mu$Jy, an
increase of a factor of 10 over its detection flux.  This is an
extreme, but not unprecedented, increase in X-ray flux.  If the X-ray
flux is higher than the gamma-ray flux, as it is with Mrk 501
(\cite{Catanese97c}), the X-ray flux requirements are even greater.
Contemporaneous observations of 1ES 2344+514 will do much to clarify
the shape of its SED for comparison with the other VHE-emitting BL Lac
objects.

No other northern hemisphere imaging Cherenkov telescope has reported
a detection or upper limit on 1ES 2344+514 at this time, though none
were observing this object during 1995.  Just as with Mrk 501, 1ES
2344+514 is not a source in the EGRET catalogs (\cite{Fichtel94};
\cite{Thompson95}; \cite{Thompson96}).  The All-Sky Monitor on the
{\sl Rossi X-ray Timing Experiment} will allow better long-term
multiwavelength monitoring of this object and continued monitoring of
this object with Cherenkov gamma-ray telescopes will hopefully confirm
this object as a VHE gamma-ray source in the near future.

\acknowledgments 

We acknowledge the technical assistance of K. Harris, T. Lappin, and
E. Roache.  We thank E. Perlman for providing an early version of his
paper on the Einstein Slew Survey BL Lac objects which enabled us to
pursue observations of 1ES 2344+514 in 1995.  We also thank the EGRET
team for providing the $>$100 MeV flux upper limits for 1ES 2344+514.
This research is supported by grants from the U. S.  Department of
Energy and by NASA, by PPARC in the UK and by Forbairt in Ireland.

\appendix

\section{Tracking mode observations}
\label{trkmode}

In the Tracking observation mode the On-source and background data are
recorded at identical times, so changes in external weather conditions
or system-wide variations in the electronics should affect both the
On-source and background regions similarly.  The Tracking mode also
allows more On-source data to be taken in a given amount of time
because no separate Off-source run is needed.  Finally, because more
events ($\sim$3 times as many) are used in the background estimate
than in the On/Off analysis, it can lead to more statistically
accurate estimates of the mean background level.  However, the
Tracking analysis requires the distribution of background events
versus $\alpha$, the orientation parameter, to be the same from run to
run in order to make accurate background estimates.  Variations in the
noise which do not affect the camera uniformly could change the
$\alpha$ distribution.  An example of such a variable noise source is
the background light distribution due to stars within the FOV which
changes with observation field and sky clarity.  Also, a PMT which is
turned off for a period of time, due to a bright star or malfunction,
affects the telescope response in a non-uniform way.  If the $\alpha$
distribution changes, the factor which converts the background events
(i.e., those with $\alpha = 20^\circ - 65^\circ$) to an On-source
($\alpha = 0^\circ - 15^\circ$) background estimate changes as well.
If this factor, called a tracking ratio, varies for different
observations, the tracking analysis is not a viable means of
estimating the background.

To investigate these concerns, we have analyzed 148 hours of data
taken on 20 non-source observation fields in the 1995/96 season and 99
hours of data taken on 19 non-source observation fields in the 1996/97
season.  All runs were taken under good weather conditions.  We
analyzed these two data sets separately because they lead to different
tracking ratios for the two seasons as shown below.  These
observations consist of Off-source runs and On-source runs for objects
which do not reveal even a marginally significant signal.  The only
other selection made on these observation fields was that they not
have any stars within the field of view which produce a current
greater than $\sim$100 $\mu$A in more than one PMT.  Stars which are
that bright, such as $\zeta$ Tauri 1$\fdg$1 from the Crab Nebula, have
a clear effect on the tracking ratio.  The non-source fields used here
have a range of mean background brightness levels due to low magnitude
stars, and the elevations at which the observations were taken range
from 20$^\circ$ to 85$^\circ$.  So, no particularly restrictive
constraints are placed on the data that we use to determine a tracking
ratio.  This is as it should be; if we must be very restrictive about
the type of field we use in a tracking ratio estimate, then the
tracking analysis loses its usefulness.

A tracking ratio of 0.292$\pm$0.004 is found for the 1995/96 data set
and the 1996/97 data set indicates a tracking ratio of
0.316$\pm$0.004.  The differences in the two tracking ratios are most
likely due to the fact that the telescope camera PMT gains were set to
different values in the two seasons, resulting in a different response
of the telescope to the Cherenkov and background light.

To determine whether the different observation fields show evidence of
inter-run, inter-source, or intra-season variability in the tracking
ratio, we applied several tests.  To investigate the stability of the
tracking ratio from run-to-run, we calculated the $\chi^2$ probability
that the tracking ratios for the individual runs were consistent with
a single tracking ratio.  For the 1995/96 data, the probability that
the individual runs are drawn from the statistical distribution of a
single tracking ratio is 0.23 and for the 1996/97 data, it is 0.16.
Both are thus consistent with arising from a single tracking ratio.

To test whether the different observation fields have different
tracking ratios, we determined the significances of the excesses or
deficits of the observations taken on the different fields and also
calculated the $\chi^2$ probability that the objects' mean tracking
ratios were drawn from a distribution with a common mean.  If the
tracking ratio varies with observation field, using a single tracking
ratio will result in a significant excess or deficit in some fields.
Among the 39 fields used in the tracking ratios for the two seasons,
only one had a statistical excess or deficit of more than 2$\sigma$.
This is expected 86\% of the time given the number of trials in this
test.  The probability that the means of the tracking ratios are drawn
from a single mean is 0.73 for the 1995/96 data and 0.44 for the
1996/97 season.  So, both tests indicate that the tracking ratios for
the individual fields are consistent with a single tracking ratio.

The final test we applied was to see if the tracking ratio changed
over the course of the season.  To do this, we calculated the
probability that the excess or deficit counts per minute for the
summed objects were consistent with statistical fluctuations about a
constant mean.  If the tracking ratio changes with time, it should
produce a consistent excess or deficit that gives a poor fit to the
constant mean hypothesis.  The $\chi^2$ probability that the rate is
constant is 0.68 for the 1995/96 data set and 0.25 for the 1996/97
season.  Thus, we have no evidence that the tracking ratio changes
significantly over the course of an observing season.

>From these tests, we conclude that the single tracking ratio can be
used for objects which do not have particularly bright stars within
the telescope's FOV.  We do, however, see a significant difference in
the tracking ratio between seasons.



\begin{deluxetable}{c}
\tablewidth{0pt}
\tablecaption{Supercuts95 \label{s95}}
\tablehead{\colhead{Gamma-ray range}}
\startdata
{\sl max1} $>$ 100 d.c.\tablenotemark{a} \nl
{\sl max2} $>$ 80 d.c. \nl
{\sl size} $>$ 400 d.c. \nl
0$\fdg$16 $<$ {\sl length} $<$ 0$\fdg$30 \nl
0$\fdg$073 $<$ {\sl width} $<$ 0$\fdg$15 \nl
0$\fdg$51 $<$ {\sl distance} $<$ 1$\fdg$10 \nl
$\alpha$ $<$ 15$^\circ$ \nl
\enddata
\tablenotetext{a}{d.c. = digital counts.  
1.0 d.c. $\approx$ 1.0 photoelectrons.}
\end{deluxetable}

\begin{deluxetable}{cccrccc}
\tablewidth{0pt}
\tablecaption{1ES 2344+514 Observation Summary  \label{data}}
\tablehead{\colhead{ } & \colhead{ } & \colhead{Exposure} & \colhead{$S$} & 
\colhead{Rate\tablenotemark{a}} & \colhead{Flux\tablenotemark{b}} & 
\colhead{Flux\tablenotemark{c}} \\
\colhead{Epoch} & \colhead{Data type} & \colhead{(hours)} & 
\colhead{($\sigma$)} & \colhead{cts/min} & \colhead{(Crab)} & 
\colhead{($\times 10^{-11}$ cm$^{-2}$ s$^{-1}$)}}
\startdata
1995/96 & On/Off & 11.5 & 2.9 & $<$0.40 & $<$0.25 & $<$2.6 \nl
	& Tracking\tablenotemark{d} & 20.5 & 5.8 & 0.25$\pm$0.04 & 
 0.16$\pm$0.03 & 1.7$\pm$0.5 \nl
1995 Dec 20 & On/Off & 1.38 & 4.2 & 0.87$\pm$0.21 & 0.55$\pm$0.13 & 
 5.8$\pm$1.9 \nl
       & Tracking & 1.85 & 6.0 & 1.00$\pm$0.17 & 0.63$\pm$0.11 & 
 6.6$\pm$1.9 \nl
1996/97 & On/Off & 17.4 & -0.4 & $<$0.15 & $<$0.09 & $<$0.93 \nl
        & Tracking & 24.9 & 0.4 & $<$0.13 & $<$0.08 & $<$0.82 \nl
\enddata
\tablenotetext{a}{The excess counts per minute,
or the 99.9\% CL upper limit on that quantity, passing Supercuts95.}
\tablenotetext{b}{The flux, or 99.9\% CL upper limit, is expressed
as a fraction of the VHE Crab Nebula flux.}
\tablenotetext{c}{The integral flux, or 99.9\% CL upper limit, is quoted
for E $>$ 350 GeV.}
\tablenotetext{d}{In this table, Tracking refers to all On-source runs, including those
taken as part of an On/Off pair.}
\end{deluxetable}
\end{document}